\documentstyle[12pt,psfig]{article}
\begin{document}
\begin{titlepage}
\centerline{\bf Individual Entanglements in a Simulated Polymer Melt}
\bigskip
\centerline{\bf E.~Ben-Naim$^1$, G.~S.~Grest$^2$, T.~A.~Witten$^1$,
A.~R.~C.~Baljon$^3$}
\medskip
\centerline{$^1$The James Franck Institute, The University of Chicago,
Chicago IL 60637}\smallskip
\centerline{$^2$Corporate Research Science Laboratory, Exxon Research and
Engineering Company,}
\centerline{Annandale, New Jersey 08801}\smallskip
\centerline{$^3$Physics Department, Johns Hopkins University, Baltimore MD}
\smallskip
\vskip .65in
\centerline{ABSTRACT}
{\noindent We examine entanglements using monomer contacts between
pairs of chains in a Brownian-dynamics simulation of a polymer melt.
A map of contact positions with respect to the contacting monomer
numbers $(i, j)$ shows clustering in small regions of $(i, j)$ which
persists in time, as expected for entanglements.  Using the
``space''-time correlation function of the aforementioned contacts, we
show that a pair of entangled chains exhibits a qualitatively
different behavior than a pair of distant chains when brought
together. Quantitatively, about 50$\%$ of the contacts between
entangled chains are persistent contacts not present in independently
moving chains.  In addition, we account for several observed scaling
properties of the contact correlation function.}

\medskip

\noindent P.A.C.S.  83.10.N, 61.25.H, 02.70.Ns
\end{titlepage}

\centerline{\bf I. Introduction}\smallskip
Since the first theories of rubber elasticity, the notion of
entanglement between polymer chains has been a central component
\cite{Doi.Edwards.book}.  Conceptually these entanglements are viewed
as discrete constraints that limit the configurations of the chains,
reduce their randomness, and thus contribute to the entropic stress in
the rubber.  These same entanglements are thought to be present in
polymer melts. However, in a melt the entanglement constraints are not
permanent.  The diffusive reptation-like motions of each chain along its
length permit chain ends to pass through a given entanglement and
release the associated constraint \cite{deGennes}.  In these theories
entanglements are viewed as discrete objects, each making a comparable
contribution to the elastic modulus.

A complementary view of entanglements is used in the tube model
\cite{Doi.Edwards.book,deGennes}.  The tube model allows one to treat
the fractional stress remaining as a function of time after a step
strain.  The magnitude of the modulus enters through the tube
diameter, and there is no notion of discrete constraints.
On the other hand, the notion of discrete constraints reappears when
the model is refined \cite{Graessley,Pearson}.  For a complete
theory it is necessary to account for the so-called tube-renewal or
constraint-release effects.  The tube consists of
confining chains, and if an end of one of these confining chains
crosses the tube, a constraint is released from the confined chain.
As a result, an increment of stored free energy of order $k_B
T\gamma^2$ is lost in a liquid under unit strain $\gamma$.

The double-reptation model of des Cloizeaux recognizes the implicit
symmetry between the confining chain and the confined chain
\cite{des.Cloizeaux}.  It treats all the elastic stress as arising
from such constraints.  This theory focuses on a particular
entanglement constraint which involves two chains.  Such an entanglement
relaxes when either end of either chain passes through it.  Tube
disengagement and tube renewal are thereby treated as parts of a
single phenomenon.

In all the above theories of polymer elasticity, the idea of discrete,
spatially localized entanglements involving pairs of chains is present
explicitly or implicitly. Yet only average properties, which ignore
the individual nature of the entanglements, are discussed in these
theories.  In this study, we outline a strategy for identifying and
characterizing {\it individual} entanglements in simulated polymer liquids.
The strategy is based on the notion that an entanglement between two
chains produces persistent contacts between them.  Gao and Weiner
\cite{Gao.Weiner} followed a similar path in a recent study. While
they were unable to detect persisting contacts, they found that low
mobility monomers tend to form clusters.  We consider a sample of
simulated chains, and our analysis of these chains provides suggestive
qualitative and quantitative evidence for entanglements. We begin by
describing the expected behavior of interchain contacts owing to
entanglements.  Next we describe a series of statistical measurements
of these contacts using simulation results.  Finally we assess these
statistics in light of our expectations.

\medskip\centerline{\bf II. Entanglements and Persistent Contacts}\smallskip

We adopt a schematic model of a polymer chain widely used for
simulations as well as for conceptual studies: a sequence of beads
connected by anharmonic springs.  The beads repel one another with a
pairwise, short-range repulsion.  We have simulated a liquid of such
chains as described in the next section.  In such simulations the
beads move randomly under small stochastic forces as well as
mechanical forces from the springs and from the bead repulsions.  Such
simulations have been used to give strong evidence for reptation
\cite{Kremer,Kremer1}.  Two beads may be said to be in contact if
their separation is smaller than a predetermined  threshold.
Two long chains in a polymer melt that have one contact
typically have many contacts.  According to the conventional theory one
expects two such chains to be {\it entangled} in a number of places
as well.  These entanglements constrain the motion of the two chains
in such a way as to increase the elastic modulus.

Entanglement constraints necessarily lead to attractive forces
between chains. As a result, an entanglement confined to a certain to
segment of a chain implies contacts within that segment.
Furthermore, an entanglement confined to a certain segment of the
chain implies contacts within that segment. The converse is not true
since a contact at a given point does not imply an entanglement
constraint there.  Thus only a fraction of the contacts between two
chains are associated with entanglements.  The others are
``incidental'' contacts having nothing to do with entanglement
constraints.

We expect incidental contacts and entanglement contacts to behave
differently in time because entanglement constraints are long-lived.
They are only released when a chain end passes through them.
Accordingly, the forces and contacts that cause the constraint must be
long-lived as well.  Incidental contacts are not subject to such
constraints and they may appear and disappear.

One way to visualize contacts between two chains is to form a
``contact map''.  This map is simply a matrix with one element for
each pair of monomers $(i,j)$ on the two chains in question.  A
contact between monomers $i$ and $j$ is indicated by $n(i,j)=1$ in the
matrix; all non-contacting elements are zero.  Thus the contact map is
a two-dimensional lattice of ones and zeros.  We imagine the chains to
be very long so that the matrix and the number of contacts are very
large.  The map then represents an irregular, diffuse cloud of
contacts.  As the system evolves in time, the cloud
changes. Individual contacts appear and disappear, and similarly, a
local cluster of contacts may translate, expand, contract, or change
shape.  A cluster consisting of incidental contacts can evaporate
completely.  On the other hand, if the contacts are caused by an
entanglement, this is impossible since it would imply the removal of
the constraint.  Therefore, the entanglement contacts are persistent.
These contacts may move but not disappear.  They may only disappear
(or appear) when the entanglement constraint does so.  According to
theory this happens when the entanglement goes to a chain end.  Thus,
entanglement contacts may only disappear when they have migrated to
the boundaries of the contact map.  Persistent contacts between two
chains may also occur without entanglement.  For example, the two
sections of the chain may be held together by other chains, such as a
third chain with a small loop through which the two examined chains
pass.  The resulting constraints may then be released if the loop is
removed.  This can happen if an end of the third chain passes through
the constrained region.

The entanglements should be observable in various ways on the contact
map.  Qualitatively an entanglement is expected to appear as a
localized cluster of contacts that maintains its average size and
density as its position moves randomly over the map.  If a sequence of
maps is stacked to make a three-dimensional cloud of points, such an
entanglement would appear as a loose, ropelike object persisting in
the time direction.  One may gain more quantitative information by
analyzing the density of contacts $n(i,j,t)$.  Since individual
contacts have little significance in themselves, it is useful to
average this density over a few monomers in $i$ and $j$.  The
resulting density $\bar n$ is a smooth function of position and time.
We expect this density field to have two contributions with
qualitatively different dynamics.  The entanglement-induced contacts
decay significantly slower than the incidental contacts.

\medskip\centerline{\bf III. Simulation}\smallskip

The molecular dynamics simulation we used is identical to that used by
Kremer and Grest \cite{Kremer,Kremer1} to study an equilibrium polymer
melt. The melt is an ensemble of 200 chains,  each containing
$N=350$ beads connected by springs.  Individual
monomers move according to the equation of motion
\begin{equation}
m\ddot{\bf r}_i=\nabla\sum_{i\ne j} U_{ij} -m\Gamma \dot{\bf r}_i
+{\bf W}_i(t).
\end{equation}
The potential $U_{ij}$ contains two contributions. The first
is a
repulsive Lennard-Jones potential
$U_{ij}=4\epsilon[(\sigma/r_{ij})^{12}-(\sigma/r_{ij})^{6}+1/4]$ for
$r_{ij}<2^{1/6}\sigma$ and 0 otherwise between
all monomers. The second is an anharmoic interaction
between bonded monomers \cite{Kremer}.  Here $\Gamma$ is the
bead friction and  ${\bf W}_i(t)$ is the random force
acting on bead $i$.  The strength of the random force is coupled to
the bead friction by the fluctuation-dissipation theorem. The
temperature $T=\epsilon/k_B$ and the density
$\rho=N/V=0.85\sigma^{-3}$.  The equations are integrated with a time
step $\Delta t=0.013\tau_0$, where $\tau_0$ is the standard of time in
Lennard-Jones units, $\tau_0=\sigma\sqrt{m/\epsilon}$.  The temporal
behavior of the melt is characterized by several time scales. For
$t<\tau_e\cong 1.8\times10^3\tau_0$ the chains have yet to encounter
the topological constraints, and the dynamics are well described by
the Rouse model \cite{Rouse}. By comparing the mean-square
displacement of a monomer in the Rouse relaxation regime with the
Doi-Edwards theory \cite{Doi.Edwards.book}, Kremer and
Grest  found that the entanglement length for this
model is
$N_e\simeq 35$ \cite{Kremer}.  Thus the chain length in the present
simulation is approximately $10N_e$. The corresponding Rouse time
scale for relaxation of chains along the tube,
$\tau_R=\tau_e(N/N_e)^2\cong1.8\times10^5\tau_0$ is comparable with
the duration of the simulation, {\it viz.}, $2.2\times 10^5\tau_0$.
The time required for the initial entanglement constraints to relax
completely, $\tau_d\sim N^3$, is beyond the simulation's temporal
range.

We recorded contacts between all pairs of chains in intervals of
$10^4\Delta t$. We defined two monomers to be in contact when their
separation $r_{ij}\le 1.5\sigma$.
We considered 20 pairs of
chains that had a large ($>100$) number of contacts at time $t=0$. The
behavior of contacts between the various pairs of chains followed the
same pattern, and in the following we discuss contacts between two
representative pairs of chains (chains 66 and 72, chains 1 and 96).
These two pairs had an average number of contacts over
the entire run of 180 and 100,
respectively.  We also considered chains that had no contacts during
the simulation. In the following section we define phantom contacts
between two distant chains.  We will present data for two such typical
pairs (chains 1 and 3, chains 1 and 4).  Contact maps are discussed in
the next section, while in section V a statistical analysis is carried
for the real contacts as well as the phantom contacts.  For
convenience, we omit the chain labels and refer to the contacts as
real contacts (66-72 and 1-96) or as phantom contacts (1-3 and 1-4).

\medskip\centerline{\bf IV. Contact Maps}\smallskip

The anticipated behavior of the contacts is nicely demonstrated by a
space-time plot of the contact map. As shown in Figure 1(a), contacts
between two monomers diffuse in space, disappear and
reappear. However, certain contacts strongly persist in space forming
``ropelike'' structures centered around two monomers.  Such structures
might indicate that the two chains are subject to a topological
constraint in the vicinity of these two monomers. To understand the
role played by entanglements we construct ``phantom contacts'', {\it
i.e.}  contacts between chains that are not subject to entanglement
constraints.  We construct these phantom contacts by identifying two
distant chains with no real contacts. We then define a phantom chain
by translating a copy of the second chain so that its center of mass
coincides with that of the first chain at time $t=0$. The contacts
between the phantom chain and the real chain are the phantom
contacts. As the simulation proceeds, the contacts between the first
chain and the translated second chain were identified, maintaining the
same translation used initially. In this way, we construct a map of
phantom contacts analogous to our map of real contacts.  Since the
phantom chain is far from the real chain, the phantom contacts cannot
be influenced by interactions between the two chains, including
entanglements. Yet according to the tube model, these contacts would
evolve in the same way as in real chains, since in this model the
chains are presumed to move independently whether they are near or far
from each other.

Figure 1(b) nicely demonstrates the qualitative difference between the
contact map of neighboring and distant chains. The phantom contact map
consists of a large ``cloud'' of contacts, in contrast with the ropes
that characterize real contact maps. This cloud suggests that phantom
contacts have weaker spatio-temporal correlations.  We also produced a
map of all $i$-$j$ contacts during the simulation, {\it i.e.}, $\sum_t
n(i,j,t)$.  The contacts between the chains under consideration are
confined to roughly three clusters that have not relaxed during the
simulation (Figure 2(a)).  Indeed, the number of clusters is
consistent with the expected number of entanglements between two
entangled chains, $\sqrt{N/N_e}\cong 3$.  This estimate can be
obtained by considering the overlap between two random walks of proper
length in three dimensions.  In comparison with real contacts, the
phantom contact map appears more diffuse (Figure 2(b)).  The above map
suggests contacts between chains as an indicator of topological
constraints, or namely, entanglements.

\bigskip\centerline{\bf V. Statistics of Contacts}\smallskip

According to the contact map, real contacts between neighboring chains
appear qualitatively different than phantom contacts between distant
ones. In this section we present a quantitative tool for
characterizing this difference. The primary feature of the contact map
is the dominance of spatio-temporal correlations. Hence, we study the
contact correlation function for both real and phantom contacts. Since
the monomer index is equivalent to the curvilinear coordinate along a
chain, and since the dynamic behavior of this quantity is given in
terms of simple scaling laws in time, we are able to predict the
spatio-temporal properties of correlations between contacts. First, we
detail the scaling predictions, and then we present numerical
verifications of these scaling laws.

The contact correlation function provides a simple, direct way to
compare the static and the dynamic properties of the chains.  As
mentioned previously, in each time frame $\tau$, contacts between two
chains, 1 and 2 for example, are given by the function
$n(i,j,\tau)$. If monomers $i$ of chain 1 and monomer $j$ of chain 2
are in contact this function equals unity, otherwise it
vanishes. The contact correlation function, $g(x,y,t)$, is defined as
\begin{equation}
g(x,y,t)=N^{-1}\sum_{\tau}\sum_{i,j}
n(i,j,\tau)n(i+x,j+y,\tau+t),
\end{equation}
where $N=\sum_{\tau}\sum_{i,j} n(i,j,\tau)$ is the total number of
contacts.  With this definition $g(0,0,0)= 1$, and in general,
$0\le g(x,y,t)\le 1$.  The correlation function has the following
physical interpretation: $g(x,y,t)$ equals the conditional probability
that monomers $x$ and $y$ are in contact at time $t$ given that at
time $t=0$, monomers 0 on both chains were in contact.

Static properties are well described by the time independent
correlation function $f(x,y)\equiv g(x,y,t=0)$.  Moreover, since
chains in the melt have the global structure of a random walk, the
asymptotic properties of this correlation function can be predicted by
simple heuristic arguments. The quantity $f(x,y)$ is the conditional
probability that monomers $x$ and $y$ are in contact, given that both
monomers 0 are in contact.  Hence, one can view the two touching
chains as a Gaussian ring of length $x+y$. The probability that such a
chain forms a ring is simply $f(x,y)=P_{x+y}(r=0)$ where
$P_N(r)=(3/2\pi Na^2)^{3/2}\exp(-3r^2/2Na^2)$ is the end-to-end
probability distribution function of a Gaussian chain with $N$ beads
and $a$ is the monomer size.
Hence, asymptotically one has
\begin{equation}
g(x,y,0)\sim\left(x+y\right)^{-3/2}.
\end{equation}
This correlation function has the advantage that it does not depend on
the spacing between successive monomers.

The above argument should hold for real contacts as well as for phantom
contacts.  In Figure 3 we show that both the real contacts and the
phantom contacts follow the scaling relation of Equation (3).
Neighboring chains are subject to excluded volume effects, and thus
they should experience an overall reduction in the number of contacts
and in the static correlation function. Indeed, the static correlation
functions in Figure 3 is larger for phantom contacts compared
with real contacts.

To examine chain dynamics such as reptation, the time dependence of
the correlation function is necessary. As for the static case, the
leading long time behavior of the correlation function can be
understood using simple arguments.  One prediction of
reptation theory is that in a melt the typical curvilinear displacement,
$x(t)$, varies algebraically in time, $x(t)\sim t^{1/\alpha}$. The
exponent $\alpha$ takes a series of values at increasing time scales
\cite{Doi.Edwards.book}.  Different time regimes are characterized by
different $\alpha$'s. For example, on the shortest time scale, the
topological constraints have not been experienced by the chain.  The
displacement is governed entirely by Rouse dynamics,
$\alpha=2$. Once the constraints are encountered, the dynamics are
slowed down considerably, and as a result $\alpha=4$. The crossover time
between these two regimes is denoted by $\tau_e$. The typical distance
traveled in time $t$ provides us with a natural time dependent length
scale.  Furthermore, we assume that the same scales dominate the
contact correlation function. Hence, the time variable $t$ is given in
terms of the time scale associated with a curvilinear displacement
$x$, $x^\alpha$.  For simplicity, we restrict attention to the reduced
correlation function, $g(x,x,t)$, whose expected scaling behavior is
\begin{equation}
g(x,x,t)\sim x^{-3/2}\Phi(z)\quad z=t/x^{\alpha}.
\end{equation}
While in the short time limit $g$ reduces to the static correlation of
Equation (1), in the long time limit, $z\gg 1$, $g$ is a
function of $t$ only. Consequently, the limiting behaviors of the
scaling function $\Phi(z)$ are
\begin{equation}
\Phi(z)/\Phi(0)\simeq\cases{ 1 &$z\ll 1$;\cr
            A z^{-3/2\alpha} &$z\gg 1$.\cr}
\end{equation}
For sufficiently long times, $t\gg x^{\alpha}$, one has
$g(x,x,t)\sim t^{-3/2\alpha}$. In short, the complete contact correlation
function is governed by a single exponent $\alpha$.

According to the scaling prediction, $x^{3/2}g(x,x,t)$ at different
times is a function of $t/x^{\alpha}$ only. We have verified this
scaling hypothesis for both real and phantom contacts (see Figure 4).
However, the optimal exponent turned out to be $\alpha=3$ in
disagreement with the above theory. The reason for this discrepancy
may be crossover effects, as the time regime where the contacts are
observed is an intermediate one between the $\alpha=2$ and
$\alpha=4$ time regimes. In fact, the temporal range we
considered was $0\le \tau\le2.8\times 10^3\tau_0$, while the crossover
between the two relevant regimes is $\tau_e\cong 1.8\times10^3\tau_0$ as
noted in Sec.~ III.  It is expected that for longer times, the
asymptotic value $\alpha=4$ will be observed.

Figure 4 suggests that the scaling function $\Phi(z)$ is a different
one for real contacts and for phantom contacts.  Furthermore, a
universal scaling function $\Phi(z)$ characterizes correlations
between phantom contacts.  This function decreases monotonically with
increasing $z$. For real contacts, the scaling function exhibits a
gentle maximum in the vicinity of $z=1$.  The strength of the maximum
was typically 1.5 for the real chains we examined (see Figure 4).  As
a consistency check, we verified that contact maps of pairs with stronger
maximas had more ropes compared with pairs that exhibited weaker
maximas.  From the definition of the scaling function (Eq.~(4)), the
position of this peak grows with time, $x\sim t^{1/\alpha}$.  This
observation suggests that persistent constraints are changing position
along the chains, and as a result, the number of delayed contacts is
increased.  In other words, the longer one waits, the further the
displacement of the constraint reaches.

This difference between real and phantom contacts can be equivalently
illustrated by examining the time dependence of the correlation
function with fixed monomer indices.  In Figure 5, $g(9,9,t)/g(9,9,0)$
is plotted versus $t$. While for phantom contacts $g$
decreases monotonically in time, there is a peak in the real contact
correlation function at $t>0$.

In Figure 4, the correlation function was rescaled by a constant
factor such that it approaches unity when $z\to0$.  The asymptotic
behavior of the scaling function is given by Eq.~(5),
$\Phi(z)/\Phi(0)\simeq A z^{-3/2\alpha}$, as $z\to\infty$.  The
asymptotic prefactor $A$ is larger for real contacts than for phantom
contacts.  The corresponding enhancement factor $A$ strongly varied
for the pairs of real chains we examined, and larger values of $A$
were found for more localized contact maps.  In other words, a
fraction of the real contacts is persistent in time. We emphasize that
all the chains we considered exhibited statistical properties similar
to the above.

\medskip\centerline{\bf VI. Discussion}\smallskip

The intriguing contact maps suggest that real contacts are
qualitatively different from phantom contacts. These differences can
be quantified using a statistical analysis of the maps.  In this
section we discuss the meaning of these differences and their
relationship to entanglement.

We can describe the differences in terms of the properties of real or
phantom contacts near a given contact.  The overall density of
contacts near a given contact is about half as large for real contacts as
for phantom contacts.  This is reflected in the smaller $g(x,x,0)$ for
real contacts at all distances $x$.  This difference is a
simple consequence of the mutual avoidance of the real chains.
Phantom monomer beads may intersect, but real beads can not.  The
intersecting configurations and their associated contacts are not
available to the the real chains, and thus fewer contacts are observed.

An opposing effect occurs when one considers {\it delayed} contacts
near an arbitrary contact.  For a given chemical separation $x$, we
compare the number of contacts at a time delay $t$ and separation $x$
relative to the number of simultaneous contacts.  For both real and
phantom chains, the number of such contacts falls off progressively
for long delays.  Two intially contacting segments of length $x$
eventually move away from each other, and their number of mutual
contacts drops off.  The number of delayed contacts is relatively
greater for real chains than for phantom chains.  Certain delayed
contacts are more numerous than simultaneous contacts for a given
separation $x$.  They remain more numerous over delay times of the
order required for a monomer to move a chemical distance $x$.  The
delayed contacts for real chains are typically 50 percent  more numerous
than for phantom chains.  In other words, contacts near a given
contact tend to persist longer in time for real chains than for
phantom chains.

These extra persistent contacts may arise from entanglement between
the two real chains.  Indeed, persistence of contacts over time was
the signature of entanglement anticipated in the Introduction.  Our
results suggest that roughly 50 percent of the contacts between two
interpenetrating chains in a melt have motion that is dictated by
entanglement constraints.  For all 20 pairs of chains that we examined,
this excess  was between 30 and 50 percent, while there was
no such fluctuation in delayed contacts for the phantom chains.  This
is a natural consequence of the discreteness of entanglements.  We
expect that some pairs of chains  have more entanglements while others
have fewer,  and those pairs with fewer entanglements should have
proportionally fewer persistent contacts.

By assuming that these persistent contacts arise from entanglements,
we can draw some implications about the entanglement contacts $n_e$
and incidental contacts $n_i$ postulated in the Introduction.  The
total number of contacts grows as the square root of the molecular
weight; the number of entanglements between two chains expected in
rubber-elasticity theory \cite{Doi.Edwards.book} grows at the same
rate.  Thus we expect $n_e/n_i$ to be independent of molecular weight.
If we identify the fractional excess of persistent contacts with
$n_e/n_i$, we infer that $n_e \cong 0.50n_i$.

There is also the hint of a characteristic size $x_0$ in the contact
correlations.  The correlation function $g(x, t)$ shows a maximum as a
function of $t$.  This maximum is weak or absent for small $x$, and it
reaches its asymptotic value for separations $x_0\cong 10$.  This
characteristic separation may reflect a characteristic size for an
entanglement.

Our analysis has led to suggestive evidence for entanglements.  The
persistent contacts we identified could well arise from other sources.
For example local regions of two chains could be held together by
other chains, thus resulting in persistent contacts.  Nevertheless,
the features we have identified appear consistent with the notion that
the persistent contacts arise from entanglements between the two
chains examined.  The number of contacts seen in the contact maps or
inferred from the decay of the correlation function is of the
appropriate magnitude.  The size scale and relative number of the
persistent contacts fits reasonably with the anticipated behavior for
discrete, local entanglements, as well.

In conclusion, these results encourage the hope that entanglements may
be identified as well-defined individual objects.  If this
identification could be made, it would lead to a deeper understanding
and better control of polymer rheology. Our analysis can be used to
distinguish entangled pairs of chains from unentangled ones.
Additionally, the contact correlation function provides a natural
probe of the dynamics of many chain systems and it will be interesting
to use this technique to study the size and the mobility of individual
entanglements. It might also prove useful to investigate properties of
contacts in other many-chain systems.

\medskip\centerline{\bf Acknowledgments}\smallskip

We thank J.~Marko and A.~Semenov for useful discussions.  This work
was supported in part by NSF under Award Number 92-08527 and by the
MRSEC Program of the National Science Foundation under Award Number
DMR-9400379.

\newpage
\medskip\centerline{\bf Figure Captions}\smallskip
\begin{itemize}

\item[{Figure 1.}] Space time representation of the contact map of
during the simulation. Shown are (a) real contacts (chains 66 and 72)
and (b) phantom contacts (chains 1 and 4).  The monomer index $j$ is
color-coded: foreground is blue, background is red.

\item[{Figure 2.}] Map of the total number of contacts between two
monomers $i$ and $j$ during the simulation.  Shown are (a) real
contacts (chains 66 and 72) and (b) phantom contacts
(chains 1 and 4).  White indicates no contacts, red a small number
of contacts and blue many contacts.

\item[{Figure 3.}] The static correlation function, $g(x,x,0)$ versus
$x$.  Real contacts (chains 66-72, 1-96, solid lines) as well as
phantom contacts (chains 1-3, 1-4, dashed lines) follow the scaling
law $g(x,x,0)\sim x^{-3/2}$. The solid line with slope $-3/2$
is shown for reference.

\item[{Figure 4.}] Scaling of the the correlation function. The
quantity $x^{3/2}g(x,x,t)$ plotted versus $z\equiv t/x^3$ for real
contacts (66-72, squares; 1-96, circles) and phantom contacts (1-3,
up triangles; 1-4, down triangles). A line with slope $-1/2$ is
shown for reference.

\item[{Figure 5.}] Time dependence of the correlation function for a
fixed monomer index. Shown is $g(9,9,t)/g(9,9,0)$ versus $t$ for real
contacts (66-72 and 1-96, solid lines) and phantom contacts (1-3 and
1-4, dashed lines).
\end{itemize}

\newpage

\begin{figure}[t]
\centerline{\psfig{figure=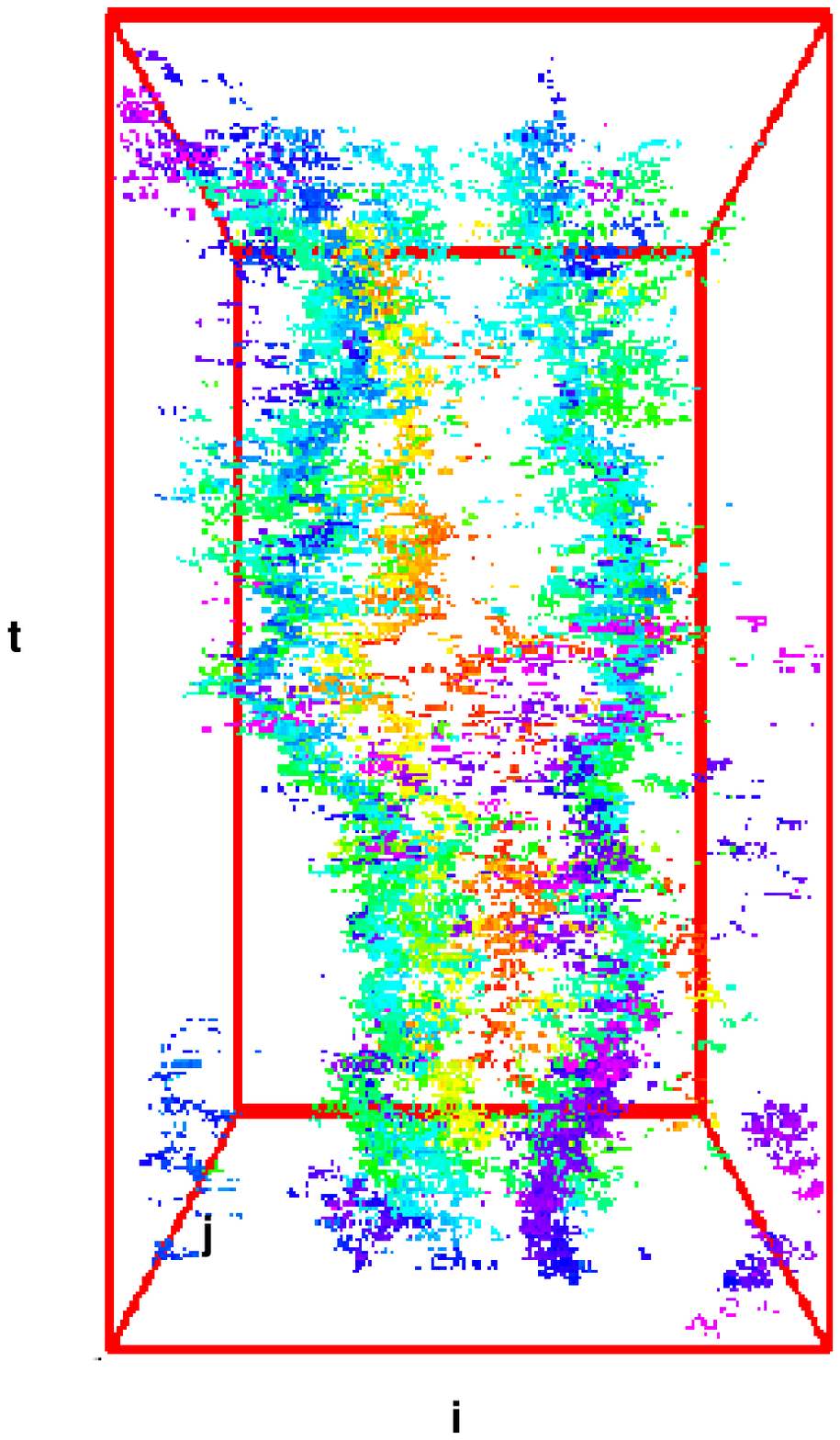}}
\end{figure}
\begin{figure}[t]
\centerline{\psfig{figure=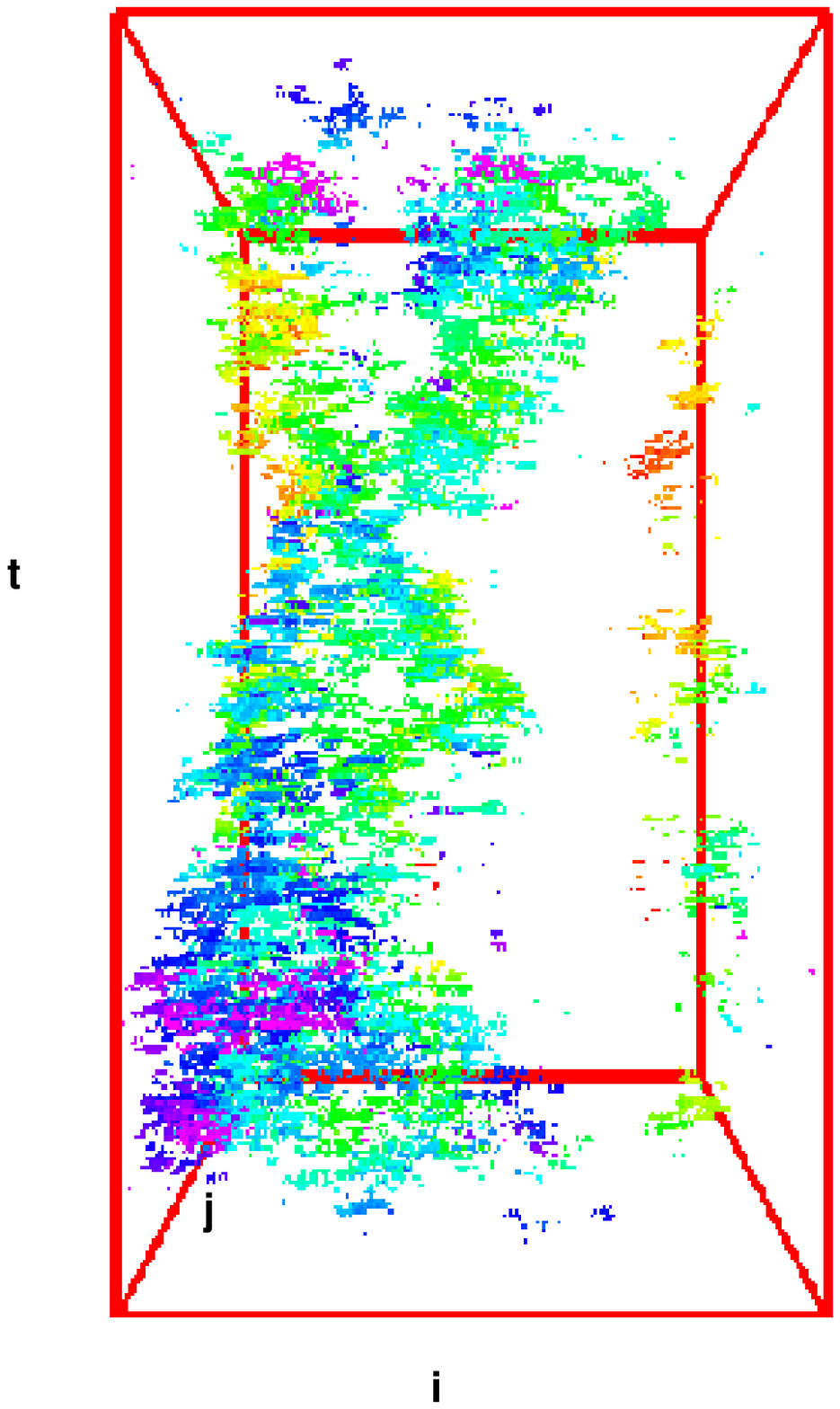}}
\end{figure}
\begin{figure}[t]
\centerline{\psfig{figure=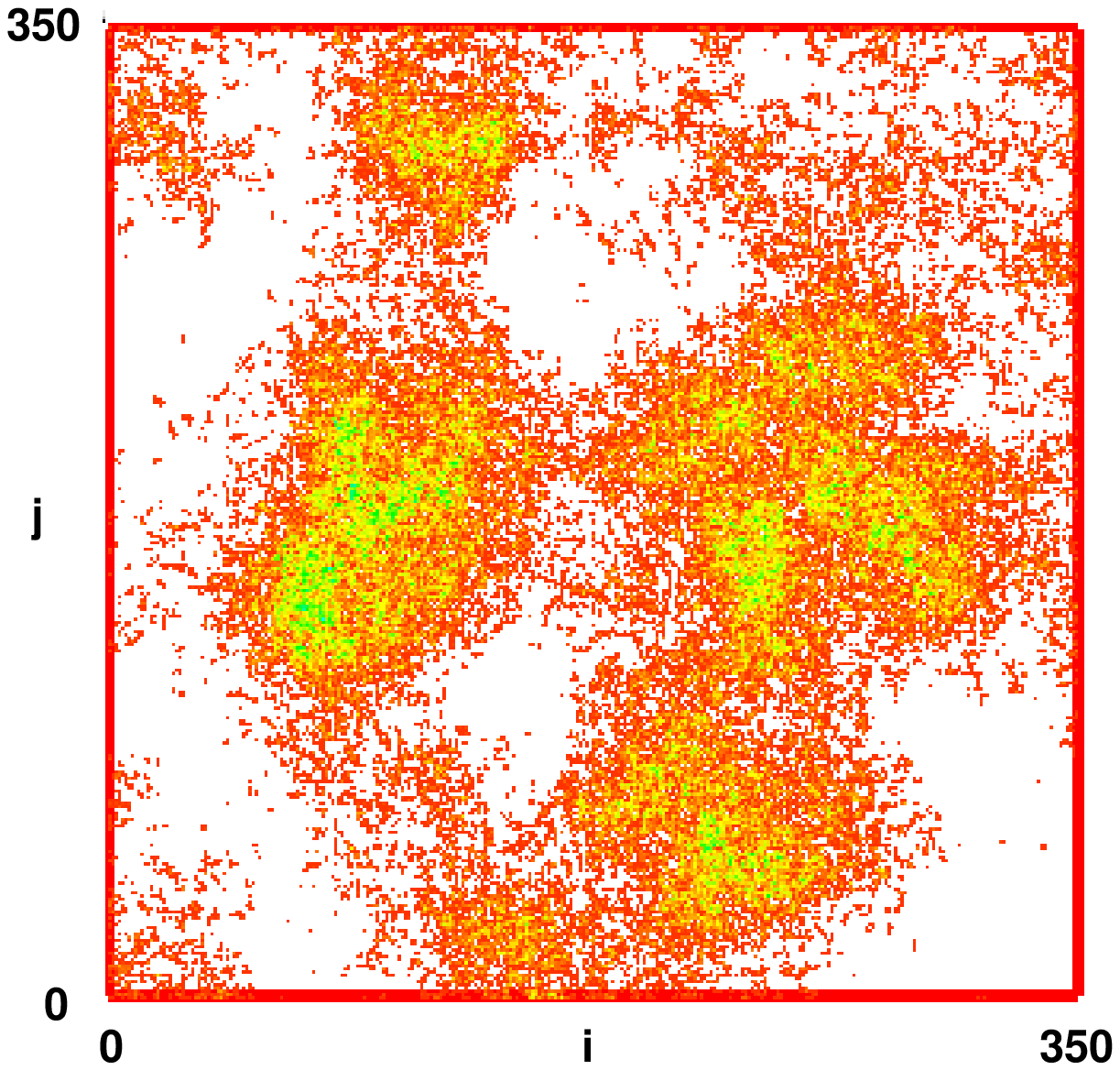}}
\end{figure}
\begin{figure}[t]
\centerline{\psfig{figure=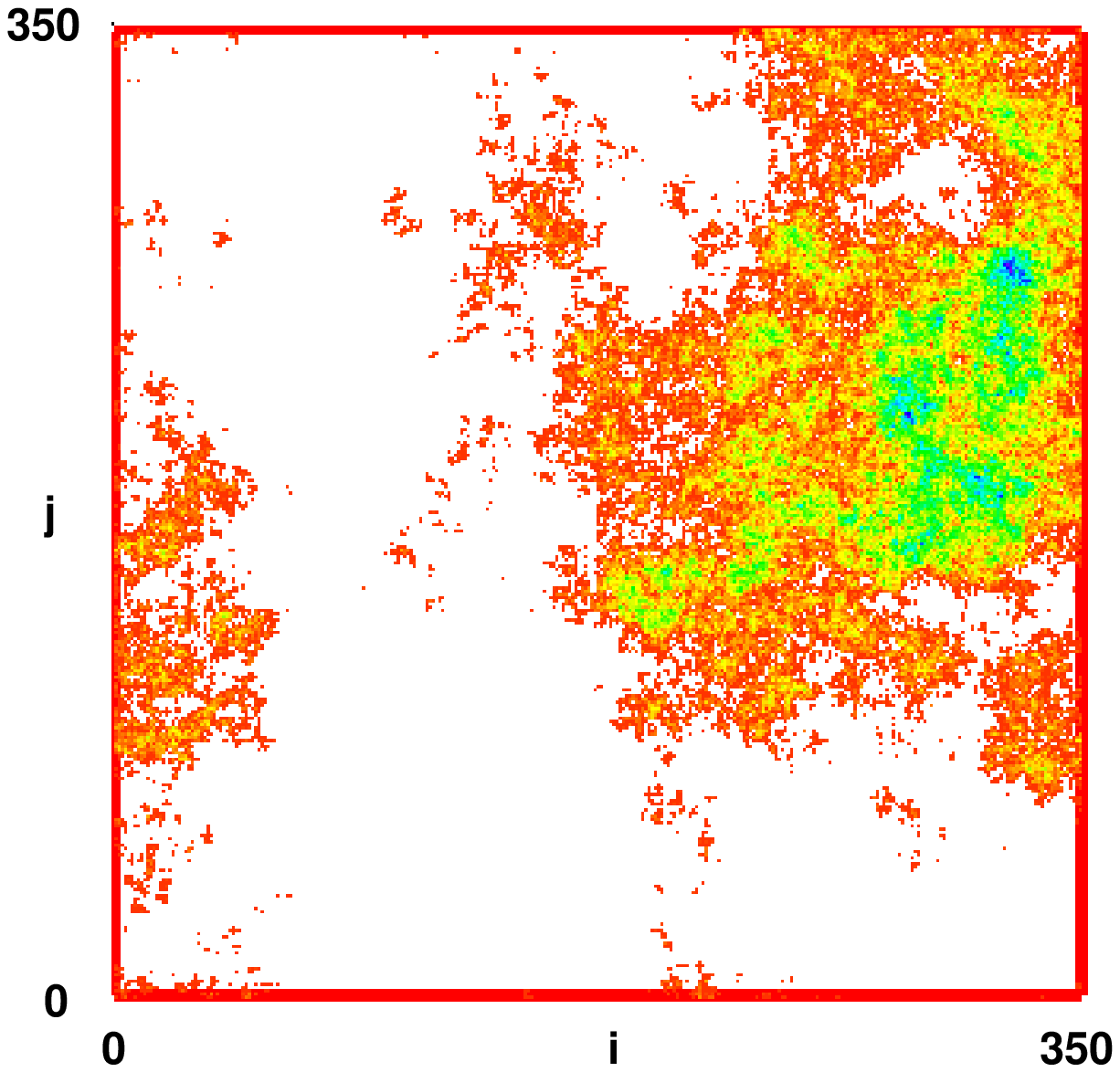}}
\end{figure}
\begin{figure}[t]
\centerline{\psfig{figure=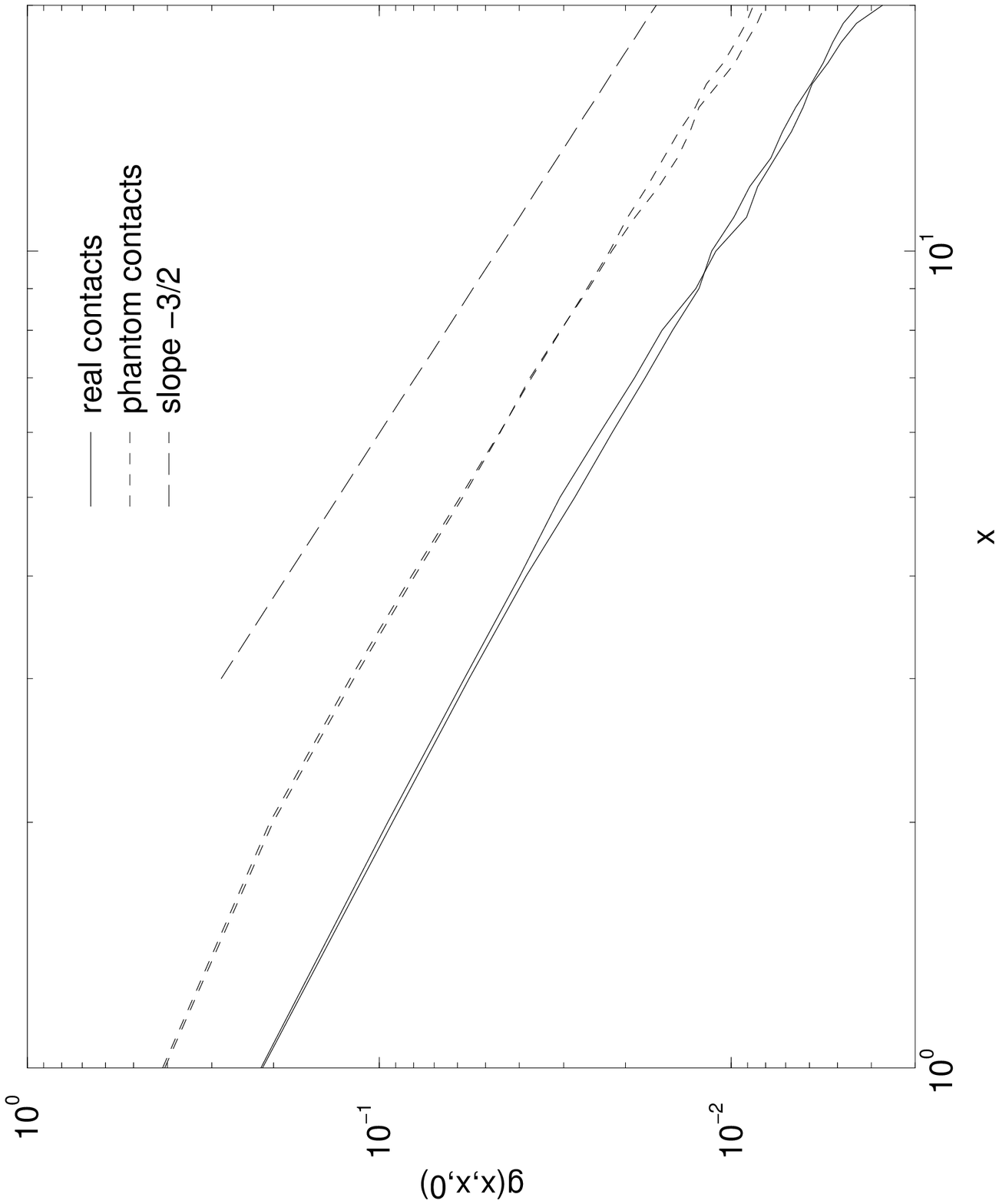}}
\end{figure}
\begin{figure}[t]
\centerline{\psfig{figure=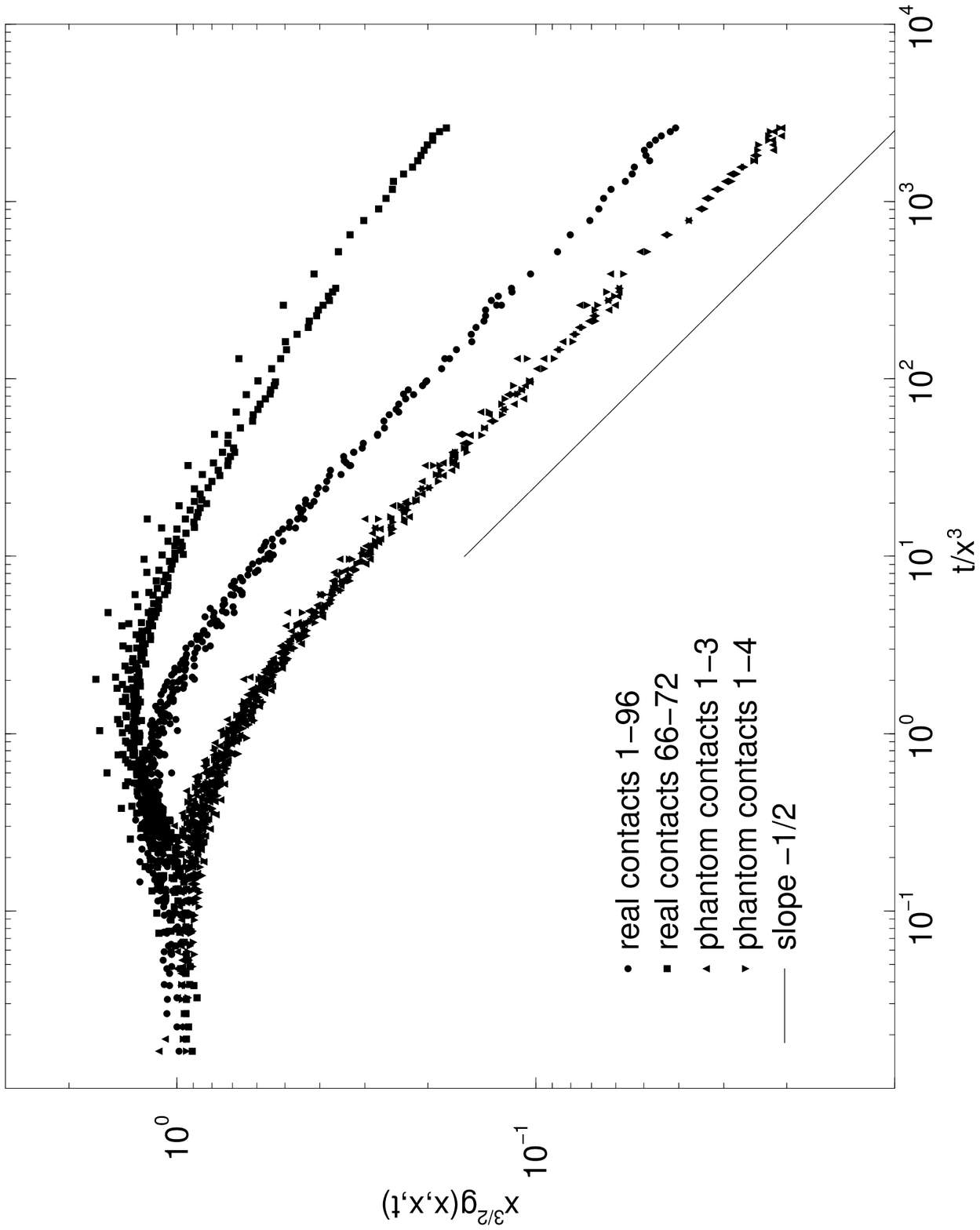}}
\end{figure}
\begin{figure}[t]
\centerline{\psfig{figure=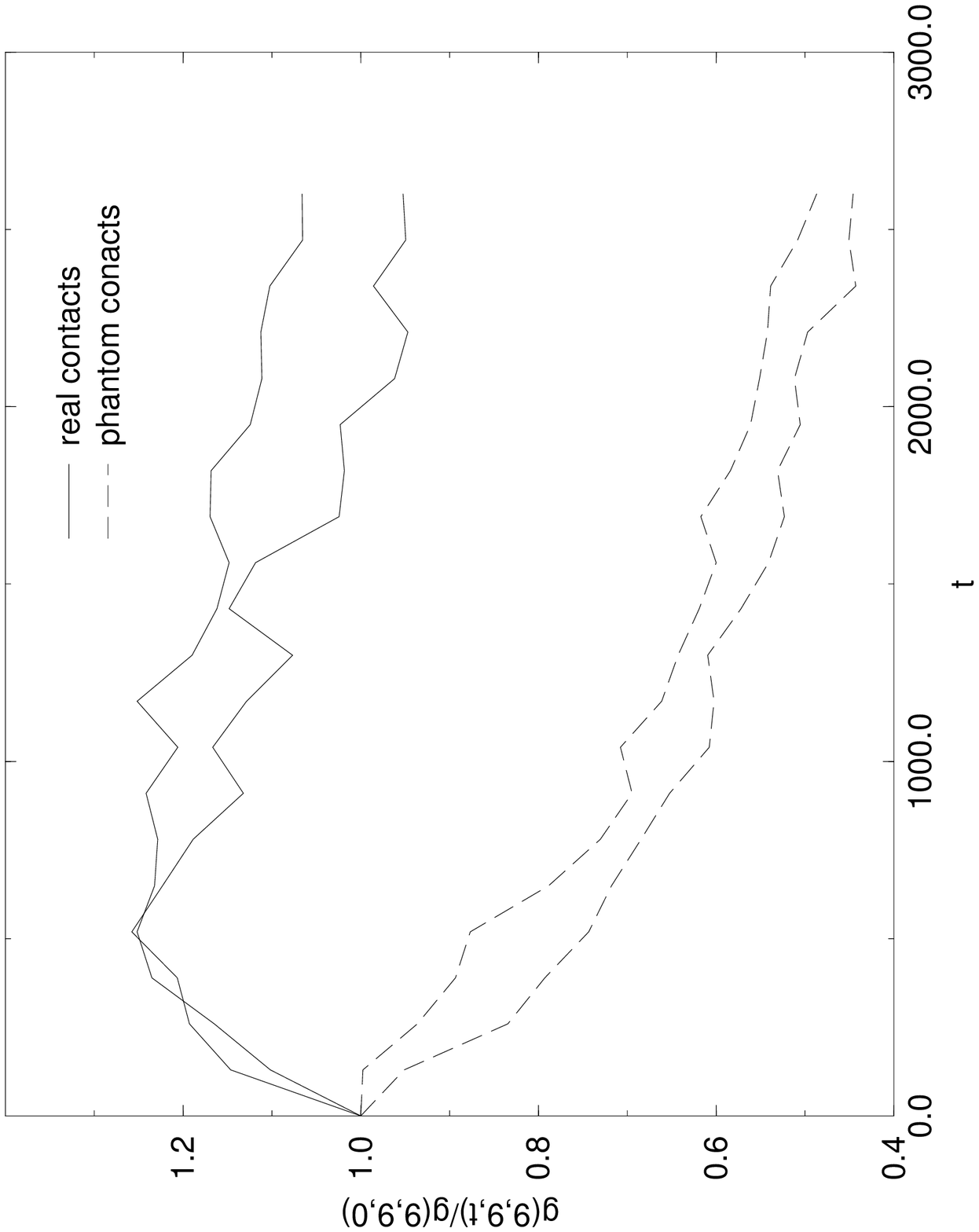}}
\end{figure}

\end{document}